\documentstyle[prb,aps,floats,epsfig,eqsecnum]{revtex}
\begin{document}
\twocolumn[\hsize\textwidth\columnwidth\hsize\csname@twocolumnfalse
\endcsname
\title{From one to two dimensions in quantum spin systems}
\author{A. Fledderjohann, K.-H. M\"utter, M.-S. Yang and M. Karbach}
\address{Physics Department, University of Wuppertal, D-42097 Wuppertal, Germany}
\date{\today}
\maketitle
%
%
%
%
\begin{abstract}
  We study the first derivative of the staggered magnetization squared
  $dm^\dag(\theta)^2/d\theta$ and the second derivative
  $d^2e_0(\theta)/d\theta^2$ of the ground state energy per site. The parameter
  $\theta$ controls the anisotropy between horizontal and vertical couplings in
  a two-dimensional (2D) spin-1/2 antiferromagnetic Heisenberg model. It is
  shown, that both derivatives diverge at $\theta=1$, where the anisotropic 2D
  model reduces to the 1D model.
\end{abstract}
\pacs{75...}
\twocolumn
]
%
\section{Introduction}
%
Antiferromagnetism looks different in the 1D and 2D Heisenberg model with
nearest-neighbor coupling of spin-1/2 matrices. The zero-temperature staggered
magnetization $m^\dag$ is known to be zero in the 1D and to be non-zero in the
2D model.\cite{KLS88,TH89,Sing89,Lian90,WOH91,Barn91,SZ92,CW93} Introducing
different strengths for the horizontal and vertical couplings $J_h\equiv
J(1+\theta)/2$ and $J_v\equiv J(1-\theta)/2$ the staggered magnetization
$m^\dag(\theta)$ has been studied as a function of the anisotropy parameter
$\theta$. It has been found in Ref.\onlinecite{AGS94} $m^\dag(\theta) \neq 0$
for $0 \leq \theta \leq \theta_0$, where $\theta_0$ is at least $\theta_0\leq
0.98$.\cite{AGS94} It has been suggested, that $m^\dag(\theta)\neq0$ in the
whole interval $0 \leq \theta < 1$.

In this paper we study the derivative ${d m^\dag(\theta)}^2/d\theta$ and the
second derivative $d^2e_0(\theta)/d\theta^2$ of the ground state energy.
The transition from one to two dimensions is performed with an interpolating
Hamiltonian ${\bf H}(\theta)$, defined on system sizes $N=k^2\pm 1,\; k=3,5,...$
with helical boundary conditions. The $\theta$-evolution of energy eigenvalues
and eigenvectors as well as expectation values of hermitian operators is
discussed in Sec. II. These general results are applied and numerically
evaluated for ${d m^\dag(\theta)}^2/d\theta$ and
$d^2e_0(\theta)/d\theta^2$ in Sec.  III and IV, respectively. Our main
result states that both derivatives diverge logarithmically with the system size
$N$ at $\theta=1$.

The origin of these divergences can be traced back to the low energy excitations
in the sector with total spin 0 of the 1D antiferromagnetic Heisenberg model.
This sector is investigated in Sec. V. In Sec. VI we discuss consequences for
the staggered magnetization in the 1D antiferromagnetic Heisenberg model with
next to nearest-neighbor couplings.
%
\section{The interpolating Hamiltonian}
%
Let us start with the 1D Heisenberg Hamiltonian with periodic boundary
conditions:
\begin{equation}
  \label{H1}
  {\bf H}_{1} \equiv \sum_{n=1}^N {\bf S}_n \cdot {\bf S}_{n+1}, \qquad 
  {\bf S}_{N+1} = {\bf S}_1,
\end{equation}
and $N=k^2\pm 1$ sites. The interpolating Hamiltonian between one and two
dimensions is defined by:
\begin{equation}
  \label{H}
  {\bf H}(\theta) \equiv \frac{1+\theta}{2} {\bf H}_{1} + 
            \frac{1-\theta}{2} {\bf Q}_k {\bf H}_{1} {{\bf Q}_k}^\dag.
\end{equation}
${\bf Q}_k$ represents a permutation operator for the sites on the ring
\begin{equation}
  \label{Qk}
  {\bf Q}_k S_n^a {\bf Q}_k^\dag = S_{(n-1)k+1}^a, \qquad a=x,y,z,
\end{equation}
which transforms the nearest-neighbor couplings in Hamiltonian (\ref{H1})  into
couplings between spins separated by $k=\sqrt{N\mp 1}$:
\begin{equation}
  \label{QH1Q}
  {\bf Q}_k {\bf H}_{1} {\bf Q}_k^\dag = 
  \sum_{n=1}^N {\bf S}_n \cdot {\bf S}_{n+k}.
\end{equation}
${\bf H}(\theta)$ defines a 2D Hamiltonian with helical boundary
conditions,\cite{HKM92,YM97} as can be seen in Fig.~1 of Ref.
\onlinecite{HKM92}. The horizontal and vertical couplings have strengths
$(1+\theta)/2$ and $(1-\theta)/2$, respectively. The parameter $\theta$ controls
the interpolation from the 1D Hamiltonian ${\bf H}(1)={\bf H}_1$ to
the 2D Hamiltonian with equal strength for horizontal and vertical
couplings
\begin{equation}
  \label{H0}
  {\bf H}_+ \equiv {\bf H}(0), 
\end{equation}
where
\begin{equation}
  \label{Hpm}
  {\bf H}_\pm \equiv \frac{1}{2}\left(
    {\bf H}_{1} \pm {\bf Q}_k  {\bf H}_{1} {\bf Q}_k^\dag
    \right).
\end{equation}
The permutation operator ${\bf Q}_k$ is essentially a rotation, which
transforms horizontal nearest-neighbor couplings into vertical
ones. One verifies the equations:
\begin{mathletters}
\begin{eqnarray}
  \label{Q2Q4}
  {\bf Q}_k^2 S_n^a {{\bf Q}_k^\dag}^2 &=& 
  \left\{ \begin{array}{ll}
          S_n^a &: k^2=N+1 \\ S_{2-n}^a &: k^2=N-1 
        \end{array}\right., \\
  {\bf Q}_k^4 S_n^a {{\bf Q}_k^\dag}^4 &=& S_n^a.
\end{eqnarray}
\end{mathletters}
Therefore, ${\bf H}_{1}$ commutes with ${\bf Q}_k^2$, but it does not commute
with ${\bf Q}_k$:
\begin{equation}
  \label{QQH1}
  [{\bf Q}_k^2,{\bf H}_{1}] = 0, \quad
  [{\bf Q}_k,{\bf H}_{1}] \neq 0.
\end{equation}
The operator ${\bf Q}_k$ transforms the interpolating Hamiltonian (\ref{H}) as
follows:
\begin{equation}
  \label{QHQ}
  {\bf Q}_k {\bf H}(\theta) {\bf Q}_k^\dag = {\bf H}(-\theta). 
\end{equation}
Note, that ${\bf Q}_k$ is a unitary operator
\begin{equation}
  \label{QQ}
  {\bf Q}_k{\bf Q}_k^\dag = {\bf 1},
\end{equation}
and for this reason ${\bf H}(\theta)$ and ${\bf H}(-\theta)$ are unitary
equivalent and possess an identical spectrum of eigenvalues:
\begin{mathletters}
\begin{eqnarray}
  \label{HPsi}
  {\bf H}(\theta)|\Psi_n(\theta)\rangle &=& E_n(\theta)|\Psi_n(\theta)\rangle,\\
  E_n(\theta) &=& E_n(-\theta), \\
   {\bf Q}_k |\Psi_n(\theta)\rangle &=& |\Psi_n(-\theta)\rangle.
\end{eqnarray}
\end{mathletters}
Let us investigate the $\theta$-evolution of the eigenvalues $E_n(\theta)$ and
the eigenvectors $|\Psi_n(\theta)\rangle$. Differentiation of Eq. (\ref{HPsi})
with respect to $\theta$ yields:
\begin{eqnarray}
  \label{dHpsi}
   \frac{d E_n(\theta)}{d \theta}\delta_{mn} &=&
   [E_m(\theta)-E_n(\theta)] \langle \Psi_m(\theta) |\Psi'_n(\theta) \rangle
   \nonumber \\ && + \langle \Psi_m(\theta)| {\bf H}_- |\Psi_n(\theta)\rangle.
\end{eqnarray}
For $n=m$ we find, that the $\theta$-evolution of the energy eigenvalues
\begin{equation}
  \label{dEn}
  \frac{d E_n(\theta)}{d \theta} = \frac{1}{2\theta} 
      \left[E_n(\theta)- 
        \langle \Psi_n(\theta) | {\bf H}(-\theta) | \Psi_n(\theta) \rangle 
      \right],
\end{equation}
is governed by the expectation values of the ${\bf Q}_k$-rotated Hamiltonian
(\ref{QHQ}). The $\theta$-evolution of the eigenvectors $|\Psi_n(\theta)\rangle$
reads
\begin{equation}
  \label{dPsi}
  \frac{d}{d\theta}|\Psi_n(\theta)\rangle = 
   -\sum_{m\neq n}
  \frac{\langle \Psi_m(\theta) | {\bf H}_- | \Psi_n(\theta) \rangle}
       {E_m(\theta) - E_n(\theta)}|\Psi_m(\theta)\rangle,
\end{equation}
which follows from (\ref{dHpsi}) for $n\neq m$. This yields for the
$\theta$-dependence of matrix elements of hermitian operators ${\bf O}(\theta)$:
\begin{eqnarray}
  \label{dO}
  \frac{d}{d\theta} 
   \langle \Psi_m(\theta) | {\bf O}(\theta) | \Psi_n(\theta) \rangle &=& 
   \langle \Psi_m(\theta) | {\bf O}'(\theta) | \Psi_n(\theta) \rangle 
   \nonumber \\ && \hspace{-35mm} -   
   \sum_{k\neq m} 
   \frac{\langle \Psi_k(\theta) | {\bf H}_- | \Psi_m(\theta) \rangle}
        {E_k(\theta)-E_m(\theta)}
        \langle\Psi_k(\theta) | {\bf O}(\theta) | \Psi_n(\theta) \rangle 
   \nonumber \\ && \hspace{-35mm} -   
   \sum_{l\neq n} 
   \frac{\langle \Psi_l(\theta) | {\bf H}_-| \Psi_n(\theta) \rangle}
        {E_l(\theta)-E_n(\theta)}
        \langle\Psi_l(\theta) | {\bf O}(\theta) | \Psi_m(\theta) \rangle.
\end{eqnarray}
In particular we get for the $\theta$-evolution of the expectation values
$\langle \Psi_m(\theta) | {\bf H}(-\theta) | \Psi_n(\theta)\rangle$, which enter
on the right-hand side of (\ref{dEn}):
\begin{eqnarray}
  \label{dHmtheta}
  \frac{d}{d\theta} 
  \langle \Psi_m(\theta) | {\bf H}(-\theta) | \Psi_m(\theta) \rangle &=&
  \nonumber \\  &&\hspace{-25mm}
  \frac{1}{2\theta} \left[
    \langle \Psi_m(\theta) | {\bf H}(-\theta) | \Psi_m(\theta)\rangle
    - E_m(\theta)
    \right] \nonumber \\  && \hspace{-25mm}  +
    \frac{1}{\theta} \sum_{n\neq m}
    \frac{\left| 
        \langle \Psi_n(\theta) | {\bf H}(-\theta) | \Psi_m(\theta) \rangle
        \right|^2}{E_n(\theta)-E_m(\theta)}.
\end{eqnarray}
It should be noted that the second term on the right-hand side of
(\ref{dHmtheta}) can be expressed in terms of the dynamical structure factor:
\begin{eqnarray}
  \label{Snw}
  S_n[\omega;{\bf H}(-\theta)] &\equiv& 
  \sum_{m\neq n} 
  \left| 
     \langle \Psi_n(\theta) | {\bf H}(-\theta) | \Psi_m(\theta) \rangle
  \right|^2 
  \nonumber \\ && \times
  \delta\biglb(\omega-[E_m(\theta)-E_n(\theta)]\bigrb),
\end{eqnarray}
associated with the transition operator ${\bf H}(-\theta)$.
%
\section{The first derivative of the staggered magnetization}
%
Equation (\ref{dO}) enables us to compute the slope of the staggered
magnetization squared:
\begin{equation}
  \label{mtheta2}
  {m^\dag(\theta)}^2 \equiv \frac{1}{N} 
  \langle \Psi_0(\theta) 
        | {\bf S}^\dag(\pi){\bf S}(\pi)
  | \Psi_0(\theta) \rangle,
\end{equation}
where
\begin{equation}
  \label{Sp}
  {\bf S}(p) \equiv \frac{1}{\sqrt{N}} \sum_{l=1}^N e^{ipl} {\bf S}_l.
\end{equation}
${m^\dag(\theta)}^2$ is known to be zero for $\theta\!=\!1$ (D=1) and nonzero for
$\theta\!=\!0$ (D=2) [ ${m^\dag(0)} \approx 0.30$ see Refs.
\onlinecite{KLS88,TH89,Sing89,Lian90,WOH91,Barn91,SZ92,CW93}]. It is expected
that ${m^\dag(\theta)} > 0$ for $0\leq\theta<1$. The derivative of the
staggered magnetization is of special interest near $\theta=1$. Using (\ref{dO})
and (\ref{mtheta2}) it follows:
\begin{eqnarray}
  \label{dmdtheta}
  \frac{{d m^\dag(\theta)}^2}{d\theta}   &=& 
       -\frac{2}{N} \sum_{m\neq 0}
       \frac{\langle \Psi_m(\theta) | {\bf H}_- | \Psi_0(\theta) \rangle}
       {\omega_m(\theta)}M_m(\theta,\pi),
       \nonumber \\ 
\end{eqnarray}
where we have introduced the notation:
\begin{equation}
  \label{Mm}
  M_m(\theta,p) \equiv 
  \langle \Psi_m(\theta) | {\bf S}(-p)\cdot{\bf S}(p) | \Psi_0(\theta) \rangle.
\end{equation}

For the computation of the right-hand side of (\ref{dmdtheta}) we need the
excitation energies $\omega_m(\theta)\equiv E_m(\theta)-E_0(\theta)$, the
transition amplitudes $\langle \Psi_m(\theta) | {\bf H}_- | \Psi_0(\theta)
\rangle$, and $M_m(\theta,\pi)$ in the total spin $S_T\!=\!0$ sector of the
interpolating Hamiltonian ${\bf H}(\theta)$. Having determined the ground state
vector $|\Psi_0(\theta)\rangle$ by means of the Lanczos algorithm, the
excitation energies and transition amplitudes can be computed via the recursion
method described in Appendix A.

Going back to the definition (\ref{Hpm}) of ${\bf H}_-$ and expressing the spin
operators ${\bf S}_l$ in terms of their Fourier transforms (\ref{Sp}), we arrive
at the following representation of the transition matrix elements:
\begin{eqnarray}
  \label{tmes}
  \langle \Psi_m(\theta) | {\bf H}_- | \Psi_0(\theta) \rangle &=& 
  \sum_p \left(e^{-ip}-e^{-ipk} \right) M_m(\theta,p).
\end{eqnarray}
It is important to note that the matrix elements $M_m(\theta,p)$ are constrained
by energy conservation for $m\neq 0$:
\begin{eqnarray}
  \label{energy_conservation}
  0 &=& \langle \Psi_m(\theta) | {\bf H}(\theta) | \Psi_0(\theta) \rangle, 
  \nonumber \\
    &=& \sum_p\left(\frac{1+\theta}{2} e^{-ip} + 
                            \frac{1-\theta}{2} e^{-ipk} \right)M_m(\theta,p). 
\end{eqnarray}
The slope of the staggered magnetization (\ref{dmdtheta}) can now be written as:
\begin{equation}
  \label{dm_dtheta_int}
  \frac{{d m^\dag(\theta)}^2}{d\theta} = 
  2\int_0^{\pi-2\pi/N} \frac{dp}{\pi} [\cos(pk) - \cos p]
  \Sigma(\theta,p,\pi,N),
\end{equation}
where we have defined:
\begin{eqnarray}
  \label{sigma}
  \Sigma(\theta,p_1,p_2,N) &\equiv& \sum_{m\neq 0} 
  \frac{M_m(\theta,p_1)M_m(\theta,p_2)}{\omega_m(\theta)}.
\end{eqnarray}
$\Sigma(\theta,p,\pi,N)$ has to obey the sum rule resulting from
(\ref{energy_conservation}):
\begin{eqnarray}
  \label{sigma_sumrule}
  \frac{2}{N}\Sigma(\theta,\pi,\pi,N) &=& 
  \int_0^{\pi-2\pi/N} \frac{d p}{\pi} \Sigma(\theta,p,\pi,N) 
  \nonumber \\ && \hspace{-5mm} \times
  [(1+\theta)\cos p + (1-\theta) \cos(pk)].
\end{eqnarray}

\begin{figure}[ht]
  \centerline{~~~~\epsfig{file=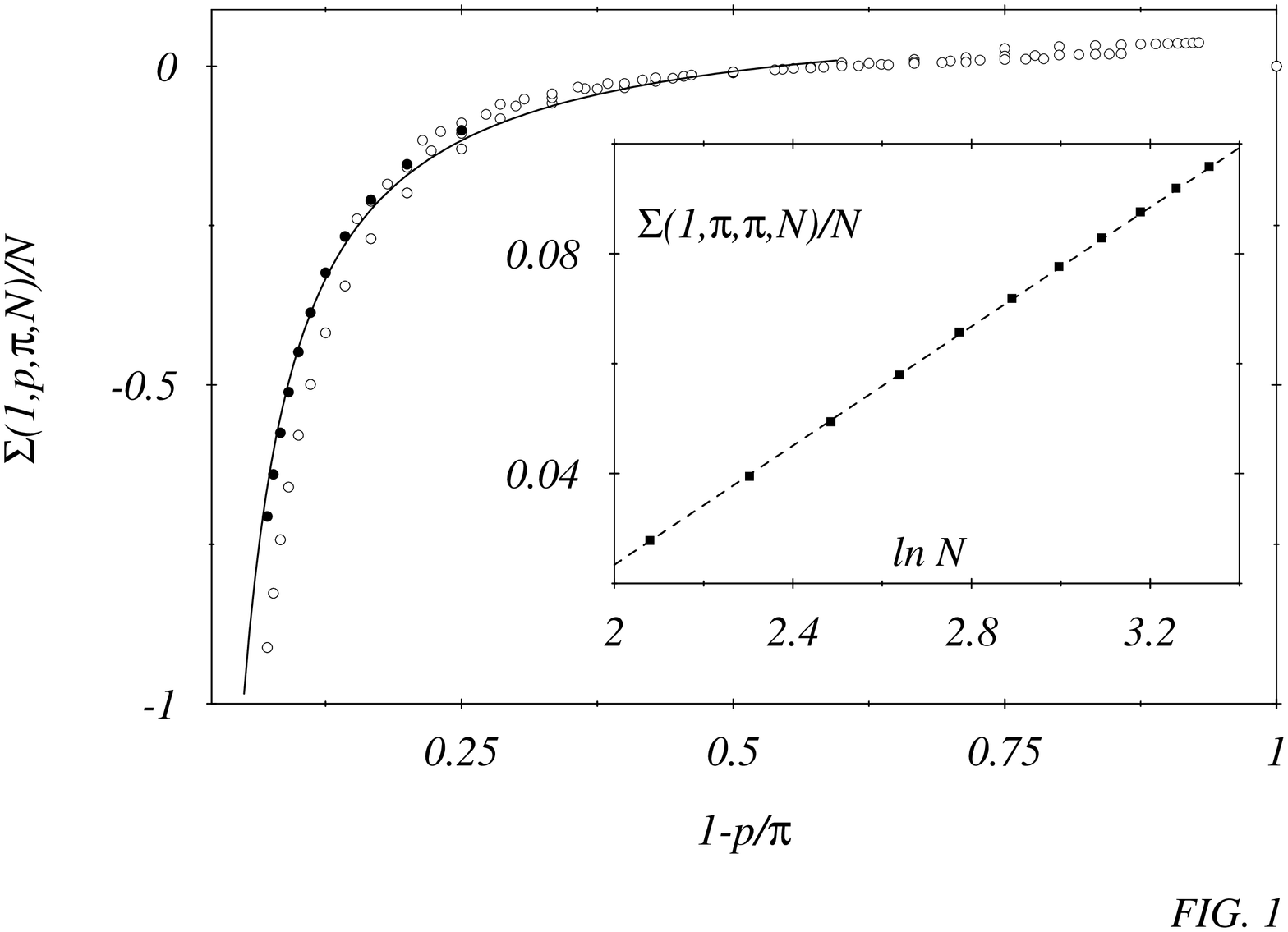,width=6.8cm,angle=-90}}
   \caption{$\Sigma(1,p,\pi,N)$ with the data of finite systems
     $N=8,10,\ldots,28,\,(\circ)$. The solid dots ($\bullet$) show the forecast
     inferred from (\ref{Sigma_N_G}). The solid line represents the prediction
     (\ref{sigma_p_pi}), whereas the dashed line in the inset represents the
     corresponding least square fit (\ref{AN0}) to $\Sigma(1,\pi,\pi,N)$
     (\ref{sigma_N_infinity}).}
   \label{fig:sigma_p_pi_N}
\end{figure}

We first discuss the situation for $\theta\!=\!1$ (D=1). As can be seen from
Fig.~\ref{fig:sigma_p_pi_N}, $\Sigma(1,p,\pi,N)$ is negative for $\pi/2\lesssim
p<\pi$, whereas $\Sigma(1,\pi,\pi,N)$ is positive by definition.

In the large $N$ limit
\begin{equation}
  \label{sigma_N_infinity}
  \frac{1}{N}\Sigma(1,\pi,\pi,N) \stackrel{N\to\infty}{\longrightarrow} 
  A\ln \frac{N}{N_0}
\end{equation}
diverges logarithmically with $N$, as is shown in the inset of
Fig.~\ref{fig:sigma_p_pi_N}.  From a least square fit we estimate:
\begin{equation}
  \label{AN0}
    A= 0.0542(5), \qquad N_0 = 1.089(5). 
\end{equation}

Owing to the sum rule (\ref{sigma_sumrule}), the divergence
(\ref{sigma_N_infinity}) is related to the singularity
\begin{equation}
  \label{sigma_p_pi}
  \Sigma(1,p,\pi,\infty) \stackrel{p\to\pi}{\longrightarrow} \frac{-A}{1-p/\pi},
\end{equation}
which we compare in Fig.~1 with finite system results for
$\Sigma(1,p,\pi,N)$. Finite-size effects are small for momentum values away from
the singularity (\ref{sigma_p_pi}). In the combined limit 
\begin{equation}
  p\to \pi, \quad N\to\infty, \quad 
  z=(1-p/\pi)N \quad \text{fixed},
\end{equation}
we try to describe the finite-size effects with a finite-size scaling ansatz:
\begin{equation}
  \label{Sigma_N_G}
  \Sigma(1,p,\pi,N) = \Sigma(1,p,\pi,\infty)G(z), 
\end{equation}
which we already used in Ref. \onlinecite{KM93,KMS94} to analyse singularities
of static structure factors. The ansatz allows to predict $\Sigma(1,p,\pi,2N)$,
for $p=\pi(1-4/N)$ and $2N=32,36,\ldots,56$. This prediction is shown in Fig.~1
by the dots ($\bullet$). These points yield a smooth extrapolation of the
numerical results for $z=4,\;N=16,18,\ldots,28$ which are already close to the
predicted behavior (\ref{sigma_p_pi}), represented by the solid curve in
Fig.~\ref{fig:sigma_p_pi_N}.  The importance of the singularity
(\ref{sigma_p_pi}) is obvious, it yields the leading contribution to the
derivative of ${m^\dag(\theta)}^2$ in the limit $N\to\infty$:
\begin{eqnarray}
  \label{dmdtheta_1}
    \left. \frac{{d m^\dag(\theta)}^2}{d\theta}\right|_{\theta=1}  
    && \stackrel{N\to\infty}{\longrightarrow}
    2A \int_{2\pi/N}^\pi \frac{d p}{p} [\cos(pk) - \cos p], 
    \nonumber \\ &&  
    \stackrel{N\to\infty}{\longrightarrow} - A\ln N.
\end{eqnarray}
We would like to stress, that the singularity (\ref{dmdtheta_1}) becomes visible
only by following the arguments presented above. In particular, the sum rule
(\ref{sigma_sumrule}) and the divergence (\ref{sigma_N_infinity}) of the
susceptibility $\Sigma(1,\pi,\pi,N)$ in the 1D case are crucial for the
derivation of (\ref{sigma_p_pi}).  

\begin{figure}[ht]
  \centerline{\epsfig{file=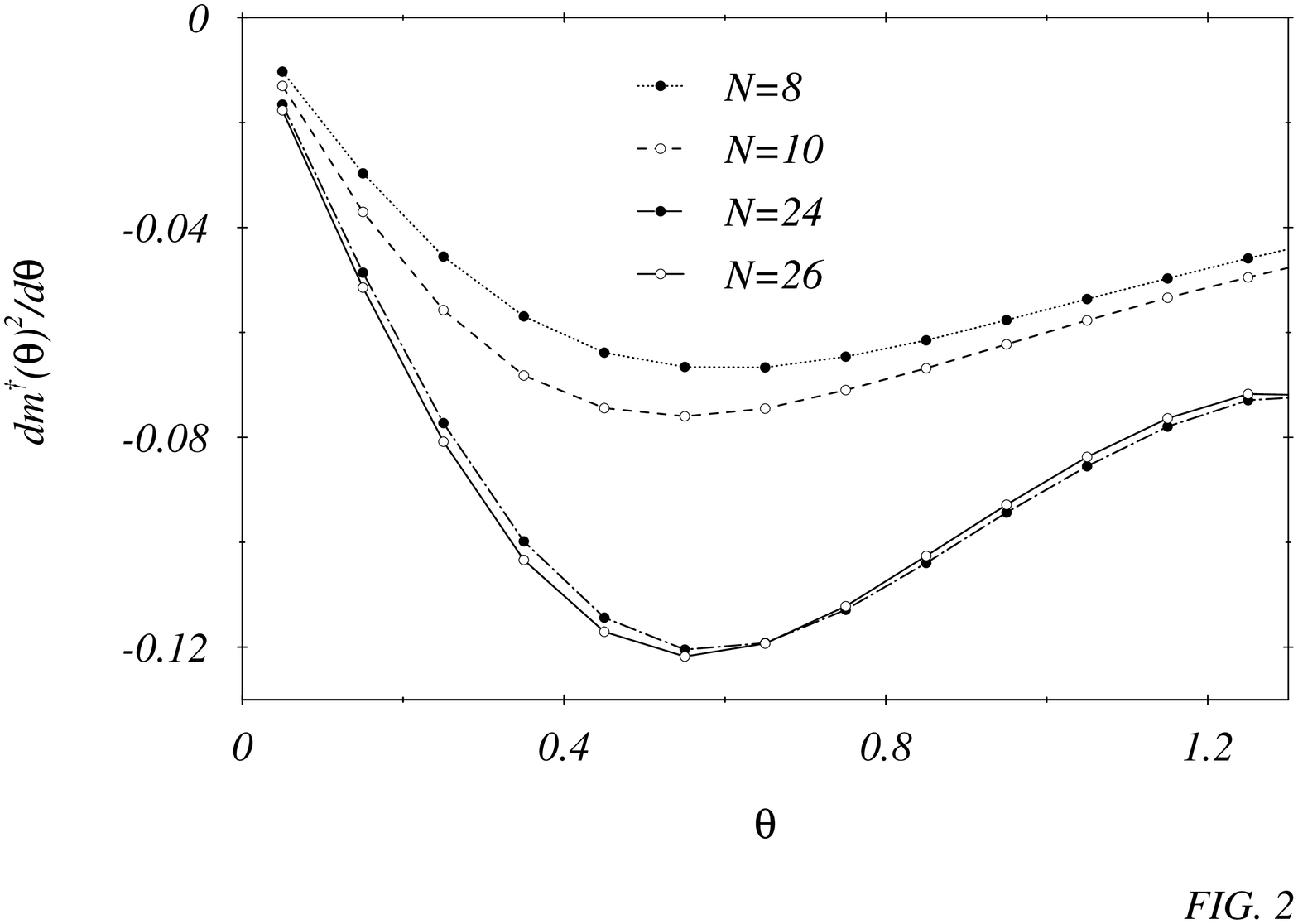,width=6.8cm,angle=-90}}
   \caption{The derivative $d{m^\dag(\theta)}^2/d\theta$ for finite systems,
     evaluated via Eq. (\ref{dmdtheta}).}
   \label{fig:dm2_dtheta}
\end{figure}

One might wonder whether the singularity of $d{m^\dag(\theta)}^2/d\theta$ at
$\theta=1$ can  be seen as well in the $\theta$-dependence. For this purpose we
have evaluated (\ref{dmdtheta}) on systems of size $N=k^2\mp 1,\; k=3,5$. The
results for $d{m^\dag(\theta)}^2/d\theta$ as a function of $\theta$ are shown in
Fig.~\ref{fig:dm2_dtheta}.

For all systems the first derivative $dm^\dag(\theta)^2/d\theta$ is negative
and decreases in the range $0<\theta<\theta_m(N)$, passes a minimum at
$\theta_m(N)$ and then increases again, by passing a turning point $\theta_t(N)$
near $\theta=1$. How to reconcile this behavior with our \text{bias} for
the thermodynamic limit of $d{m^\dag(\theta)}^2/d\theta$? We expect a decrease
for $0<\theta<1$, a singularity
\begin{equation}
  \label{scenario1}
    \frac{{d m^\dag(\theta)}^2}{d\theta}
   \stackrel{\theta\to 1^-}{\longrightarrow} -\infty,
\end{equation}
if we approach the 1D limit from the left and 
\begin{equation}
  \label{scenario2}
    \frac{{d m^\dag(\theta)}^2}{d\theta} = 0
\end{equation}
for $\theta > 1$. For $\theta>1$ the horizontal and vertical couplings in
(\ref{H}) are antiferromagnetic and ferromagnetic, respectively. In this
situation, the staggered magnetization is supposed to vanish.

If the scenario (\ref{scenario1}) and (\ref{scenario2}) is correct, there should
be a sharp discontinuity in $dm^\dag(\theta)^2/d\theta$ when we pass the
transition point $\theta=1$. For the finite systems -- presented in
Fig.~\ref{fig:dm2_dtheta} -- this sharp discontinuity is washed out. The
transition region starts already at the minimum $\theta_m(N)<1$ and extends far
beyond the transition point $\theta=1$. Therefore, we expect that for increasing
system size the minimum position $\theta_m(N)$ converges to the transition point
$\theta=1$:
\begin{equation}
  \theta_m(N) \stackrel{N\to\infty}{\longrightarrow} 1,
\end{equation}
whereas the minimum value diverges:
\begin{equation}
    \left. \frac{{d m^\dag(\theta)}^2}{d\theta}\right|_{\theta=\theta_m(N)}
   \stackrel{\theta\to 1^-}{\longrightarrow} -\infty.
\end{equation}
Comparing in Fig.~\ref{fig:dm2_dtheta} the results for $N=8,10,24,26$, we see
already that the slope of $dm^\dag(\theta)^2/d\theta$ becomes steeper for
$\theta>\theta_m(N)$. This feature could be interpreted as a signature for the
emergence of the discontinuity (\ref{scenario1}) and (\ref{scenario2}) in the
thermodynamic limit. However, it certainly requires larger systems to observe
this process.
%
\section{The second derivative of the ground state energy}
%
According to (\ref{dEn}) the first derivative of the ground state energy per
site $e_0(\theta)$ at $\theta=1$,
\begin{equation}
  \label{dE01}
  \left. \frac{d e_0(\theta)}{d\theta}\right|_{\theta=1} = \frac{1}{2} e_0(1),
\end{equation}
is given by the ground state energy $e_0(1)=-(\ln 2-1/4)$ itself.\cite{Hult38}
Here we have used the fact that the ground state expectation
value\cite{SFS89,AGSZ89}
\[
  \frac{1}{N} \langle \Psi _0 | {\bf H}(-1) | \Psi_0 \rangle = 
  \langle \Psi _0 | {\bf S}_0 {\bf S}_k | \Psi_0 \rangle 
  \stackrel{k\to\infty}{\longrightarrow} -a\frac{\sqrt{\ln k}}{k},
  \nonumber
\]
vanishes in the limit $k=\sqrt{N\mp 1} \to \infty $. Combining (\ref{dEn}) and
(\ref{dHmtheta}), we find for the second derivative:
\begin{equation}
  \label{d2e0}
  \frac{d^2e_0(\theta)}{d\theta^2} = -\frac{2}{N}
  \sum_{m\neq 0} 
  \frac{|\langle \Psi_m(\theta) | {\bf H}_- | \Psi_0(\theta) \rangle|^2}
  {\omega_m(\theta)}.
\end{equation}
The Fourier decomposition (\ref{tmes}) of the transition matrix elements
$\langle \Psi_m(\theta) | {\bf H}_- | \Psi_0(\theta) \rangle $ leads to the
representation:
\begin{eqnarray}
  \label{d2e0_int}  
  \frac{d^2e_0(\theta)}{d\theta^2} &=& -4
  \int_0^{\pi-2\pi/N} \frac{dp}{\pi}[\cos p -\cos (pk)]^2 \Sigma (\theta,p,p,N)
  \nonumber \\ &-& 
  4N\int_0^{\pi-2\pi/N}\int_0^{p-2\pi/N} \frac{dp}{\pi} \frac{dp'}{\pi} 
  \nonumber \\ && \hspace{-10mm} \times 
  [\cos p -\cos(pk)][\cos p' -\cos(p'k)] \Sigma (\theta,p,p',N).
  \nonumber \\
\end{eqnarray}
$\Sigma (\theta,p_1,p_2,N)$ is defined in (\ref{sigma}) and numerical results
for $\theta=1$ and $N=24$ are given in Table \ref{T2}.  Note, that
$\Sigma(\theta=1,p_1,p_2,N)<0$ for $\pi/2 \lesssim p_2 < p_1 < \pi$, whereas
$\Sigma(\theta=1,p,p,N)>0$ by definition.

\squeezetable
\begin{table}[htb]
 \caption{$\Sigma (\theta,p_1,p_2,N)$ [Eq. (\ref{sigma})] for $N=24$ and
   $p_1,p_2 = 2\pi l/N,\, l=4,5,\ldots,12$.}\label{T2} 
  \begin{tabular}{l|rrrrrrrrr}
  l & 4 & 5 & 6 & 7 & 8 & 9 & 10 & 11 & 12 \\ \hline
  4 & \textbf{0.005}&-0.001&-0.001&-0.002&-0.006& 0.000& 0.002& 0.003& 0.002 \\
  5 &-0.001& \textbf{0.007}&-0.002&-0.006&-0.001& 0.001& 0.001& 0.001&-0.001 \\ 
  6 &-0.001&-0.002& \textbf{0.006}&-0.001& 0.000& 0.001& 0.001& 0.000&-0.005 \\ 
  7 &-0.002&-0.006&-0.001& \textbf{0.018}&-0.000&-0.000&-0.001&-0.002&-0.011 \\ 
  8 &-0.006&-0.001& 0.000&-0.000& \textbf{0.030}&-0.002&-0.003&-0.005&-0.022 \\ 
  9 & 0.000& 0.001& 0.001&-0.000&-0.002& \textbf{0.050}&-0.007&-0.012&-0.045 \\ 
 10 & 0.002& 0.001& 0.001&-0.000&-0.003&-0.007& \textbf{0.096}&-0.029&-0.106 \\ 
 11 & 0.003& 0.001& 0.000&-0.002&-0.005&-0.012&-0.029& \textbf{0.231}&-0.371 \\ 
 12 & 0.002&-0.001&-0.005&-0.011&-0.022&-0.045&-0.106&-0.371& \textbf{1.051} \\ 
  \end{tabular}
\end{table}

$\Sigma(\theta,p,p',N)$ and $\Sigma (\theta,p,\pi,N)$ are related via the energy
conservation relation (\ref{energy_conservation}):
\begin{eqnarray}\label{sumrule_sigmap1p2}
  && [(1+\theta)\cos p + (1-\theta)\cos (pk)]\Sigma (\theta,p,p,N)
    \nonumber \\ && \hspace{10mm} - 
   \Sigma(\theta,p,\pi,N) = \Delta(\theta,p,N),
\end{eqnarray}
where
\begin{eqnarray}\label{sumrule_sigmap1p2_2}
\Delta(\theta,p,N)  &\equiv& -\frac{N}{2\pi}
  \left(
  \int_0^{p-2\pi/N}dp'+ 
  \int_{p+2\pi/N}^{\pi-2\pi/N}dp'
  \right)
  \nonumber \\ && \hspace{-15mm} \times 
  [(1+\theta)\cos p' + (1-\theta) \cos (p'k)]\Sigma (\theta,p,p',N).
  \nonumber \\ 
\end{eqnarray}

\begin{figure}[ht]
  \centerline{~~\epsfig{file=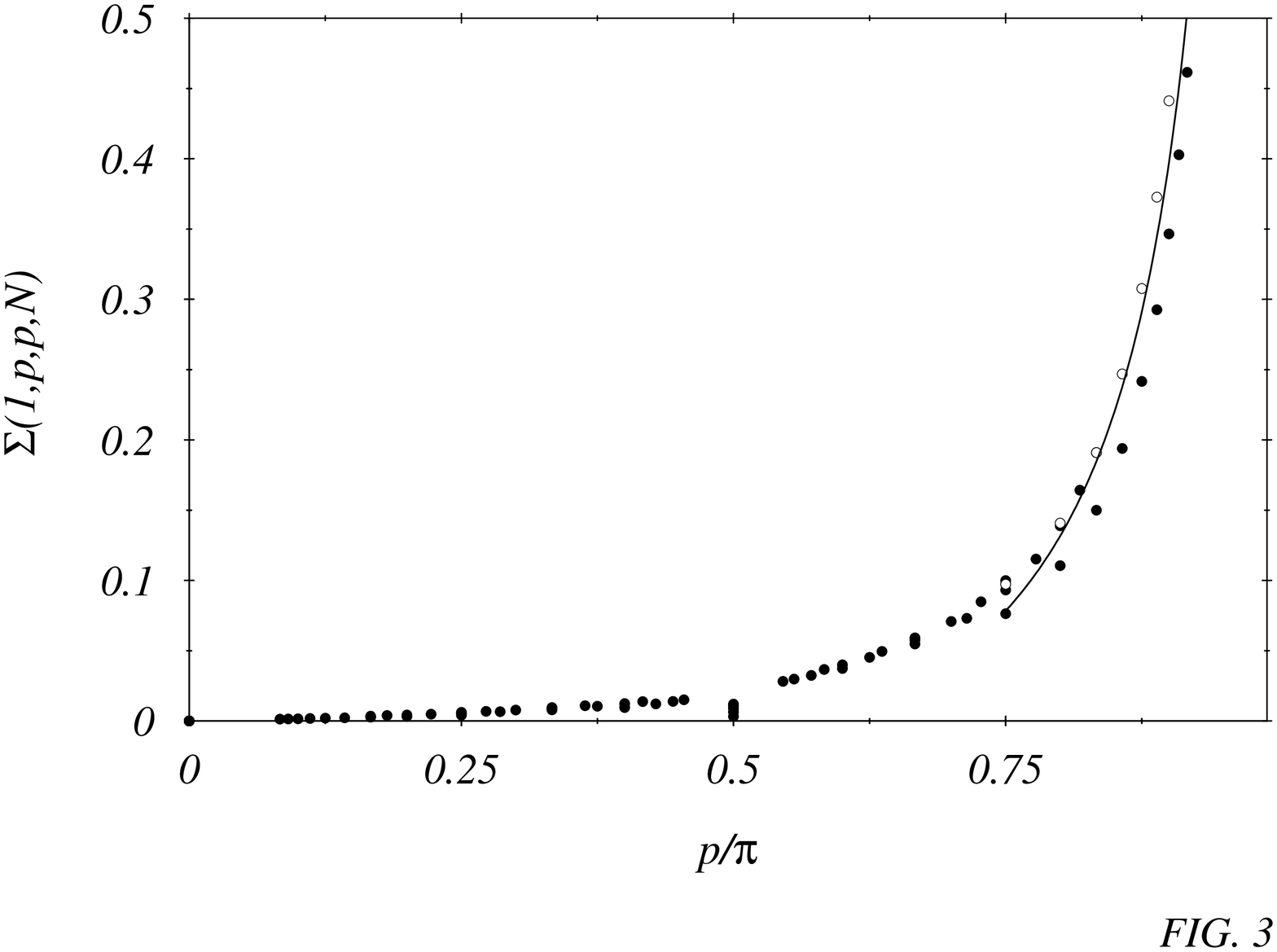,width=6.8cm,angle=-90}}
   \caption{$\Sigma(1,p,p,N)$ for finite systems
     $N=8,10,\ldots,28\,(\bullet)$ and the predicted pole (\ref{sigma_p1_p1}) in
     conjunction with (\ref{B}) for $\Sigma(1,p,p,\infty)$ (solid line).}
   \label{fig:sigma_p_p_N}
\end{figure}

The sum rule (\ref{sumrule_sigmap1p2}) for $\theta=1$ yields important
constraints for $p\to\pi$, where $\Sigma(1,p,\pi,\infty)$ develops the pole
(\ref{sigma_p_pi}) depicted in Fig.~\ref{fig:sigma_p_pi_N}. This pole has to be
compensated by a corresponding singularity in:
\begin{equation}
  \label{sigma_p1_p1}
  \Sigma(1,p,p,\infty) \stackrel{p\to\pi}{\longrightarrow} 
  \frac{B}{1-p/\pi},
\end{equation}
which is indeed observable in the numerical data shown in
Fig.~\ref{fig:sigma_p_p_N}. The finite-size effects can be treated again with a
finite-size scaling ansatz of the type (\ref{Sigma_N_G}). The open symbols
($\circ$) represent the prediction of this ansatz. The solid curve is a pole fit
of the type (\ref{sigma_p1_p1}) with:
\begin{equation}\label{B}
  B = 0.053(10).
\end{equation}

The pole for $p\to\pi$ appears as well on the left-hand side
(\ref{sumrule_sigmap1p2}) of the sum rule:
\begin{equation}
  \label{A2B}
  \Delta(\theta=1,p,N) \stackrel{p\to\pi}{\longrightarrow} 
  \frac{A-2B}{1-p/\pi}.
\end{equation}
The right-hand side of (\ref{sumrule_sigmap1p2_2}) tells us, how
$\Sigma(\theta\!=\!1,p,p',N)$ behaves for $p,p'\to\pi$ in order to generate
the pole term (\ref{A2B}) on the right-hand side of (\ref{sumrule_sigmap1p2}):
\begin{equation}
  \label{NlnNA2B}
  N\ln N \Sigma(1,p,p',N) \stackrel{p\neq p'\to\pi}{\longrightarrow}
  \frac{A-2B}{(1-p/\pi)(1-p'/\pi)}.
\end{equation}

\begin{figure}[ht]
  \centerline{~~\epsfig{file=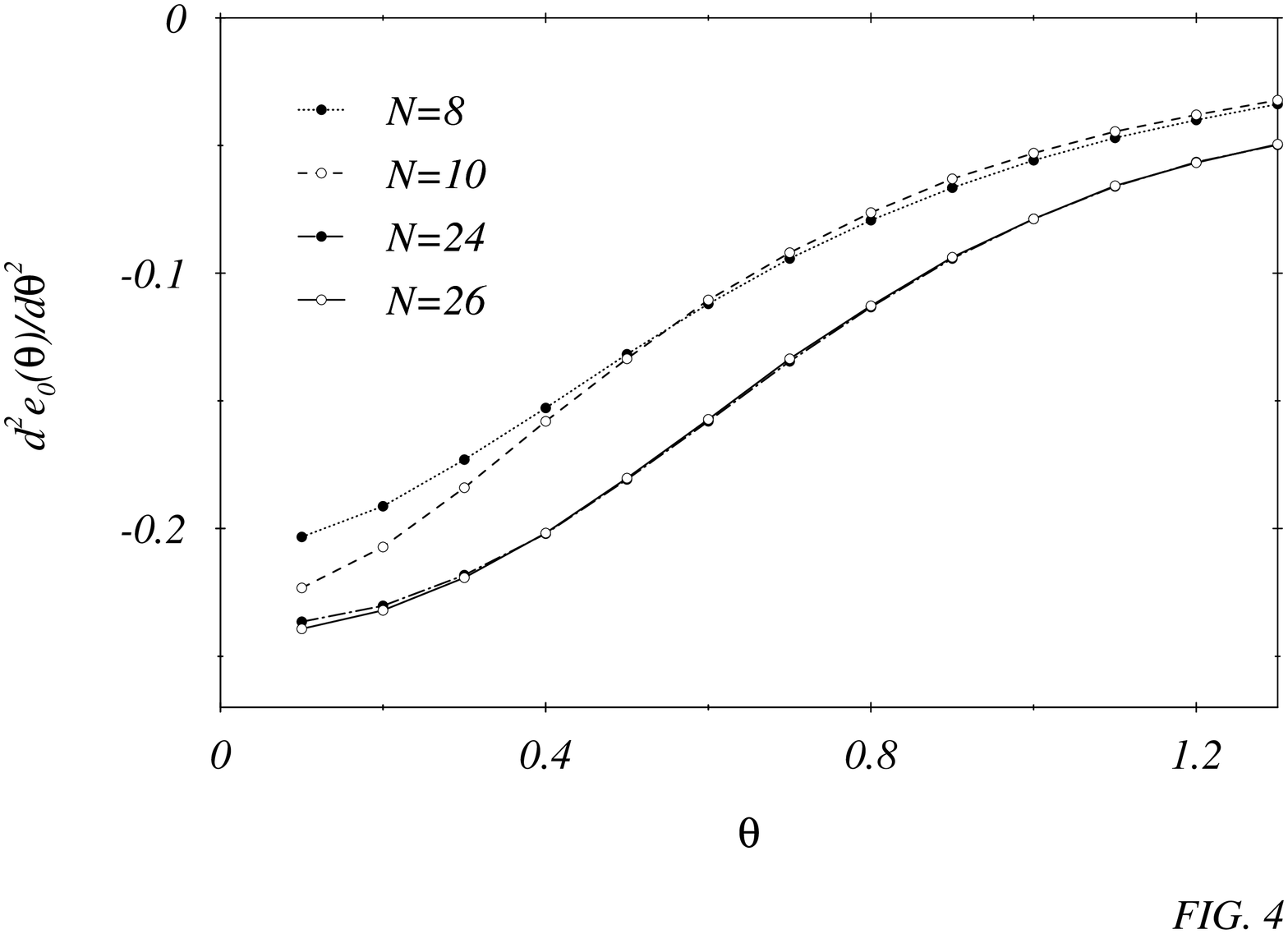,width=6.8cm,angle=-90}}
  \caption{The second derivative of the ground state energy   
  $e_0(\theta)$ versus $\theta$ as it follows from Eq. (\ref{d2e0}).}
   \label{fig:d2e_dt_t}
\end{figure}

Insertion of the leading singularities (\ref{sigma_p1_p1}) and (\ref{NlnNA2B})
into the right-hand side of (\ref{d2e0_int}) yields
\begin{equation}
  \label{d2e0_theta1}
  \left.\frac{d^2e_0(\theta)}{d\theta}\right|_{\theta=1} 
  \stackrel{N\to\infty}{\longrightarrow} -\frac{1}{2}(A+2B)\ln N.
\end{equation}
In Fig.~\ref{fig:d2e_dt_t} we present the $\theta$-dependence of the second
derivative $d^2e_0(\theta)/d\theta^2$ on finite systems $N=8,10,24,26$, as it
follows from (\ref{d2e0}).
There is no signature in the finite system results, which give a hint to the
singularity (\ref{d2e0_theta1}). A comparison of (\ref{d2e0}) with a numerical
differentiation of the ground state energy yields excellent agreement.

%
\section{Low energy excitations in the spin 0 sector of the
  1D antiferromagnetic Heisenberg model}
\begin{figure}[ht]
  \centerline{\epsfig{file=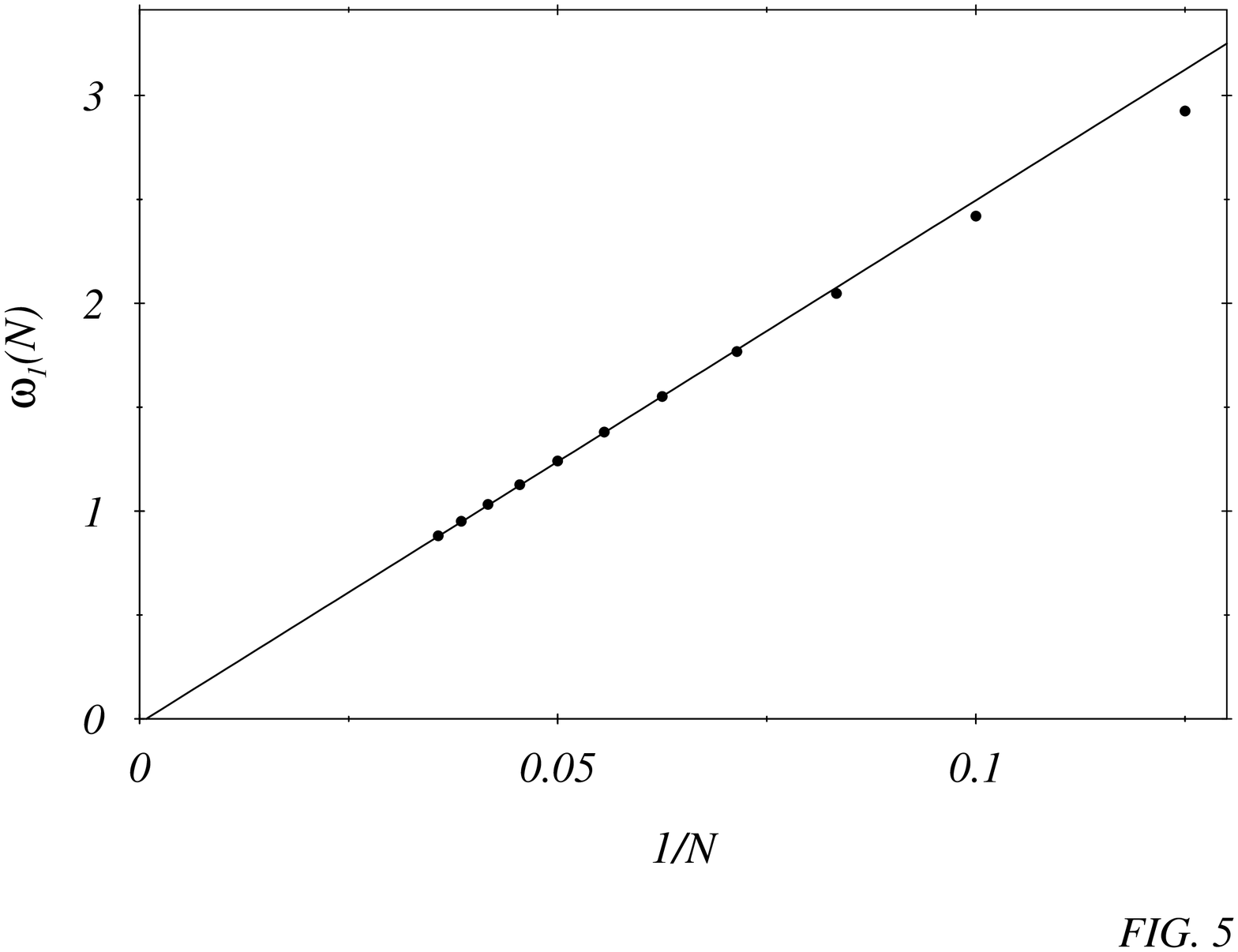,width=6.8cm,angle=-90}}
   \caption{The excitation gap $\omega_1(N)$ for $N=8,10,\ldots,28$. The solid
     line represents the fit (\ref{w1}). }
   \label{fig:w1_N}
\end{figure}

%
The singular behavior of the first derivative of the staggered magnetization and
of the second derivative of the ground state energy per site can be traced back
to the divergence (\ref{sigma_N_infinity}) of the quantity:
\begin{equation}
  \label{susz_pi_N}
  \Sigma(\theta=1,\pi,\pi,N) = \int_{\omega_1(N)}^\infty \frac{d \omega}{\omega}
  S(\omega,\pi,N),
\end{equation}
which again can be viewed as the susceptibility of the dynamic structure
factor
\begin{equation}
  \label{S_w_p_N}
  S(\omega,p,N) \equiv \sum_{n\neq 0} M_n(1,p)^2 \delta(\omega-\omega_n).
\end{equation}
The operators ${\bf S}(-p)\cdot {\bf S}(p)$, which enter in the
definition (\ref{Mm}), only allow for transitions with $\Delta S = 0$ and
$\Delta p = 0$.  The gap $\omega_1(N)$ in the sector with quantum numbers of the
ground state vanishes in the thermodynamic limit
\begin{equation}
  \label{w1}
  N \omega_1 \stackrel{N\to \infty}{\longrightarrow} \Omega_1 = 25.15(5),
\end{equation}
as is shown in Fig.~\ref{fig:w1_N}. 

\begin{figure}[ht]
  \centerline{\epsfig{file=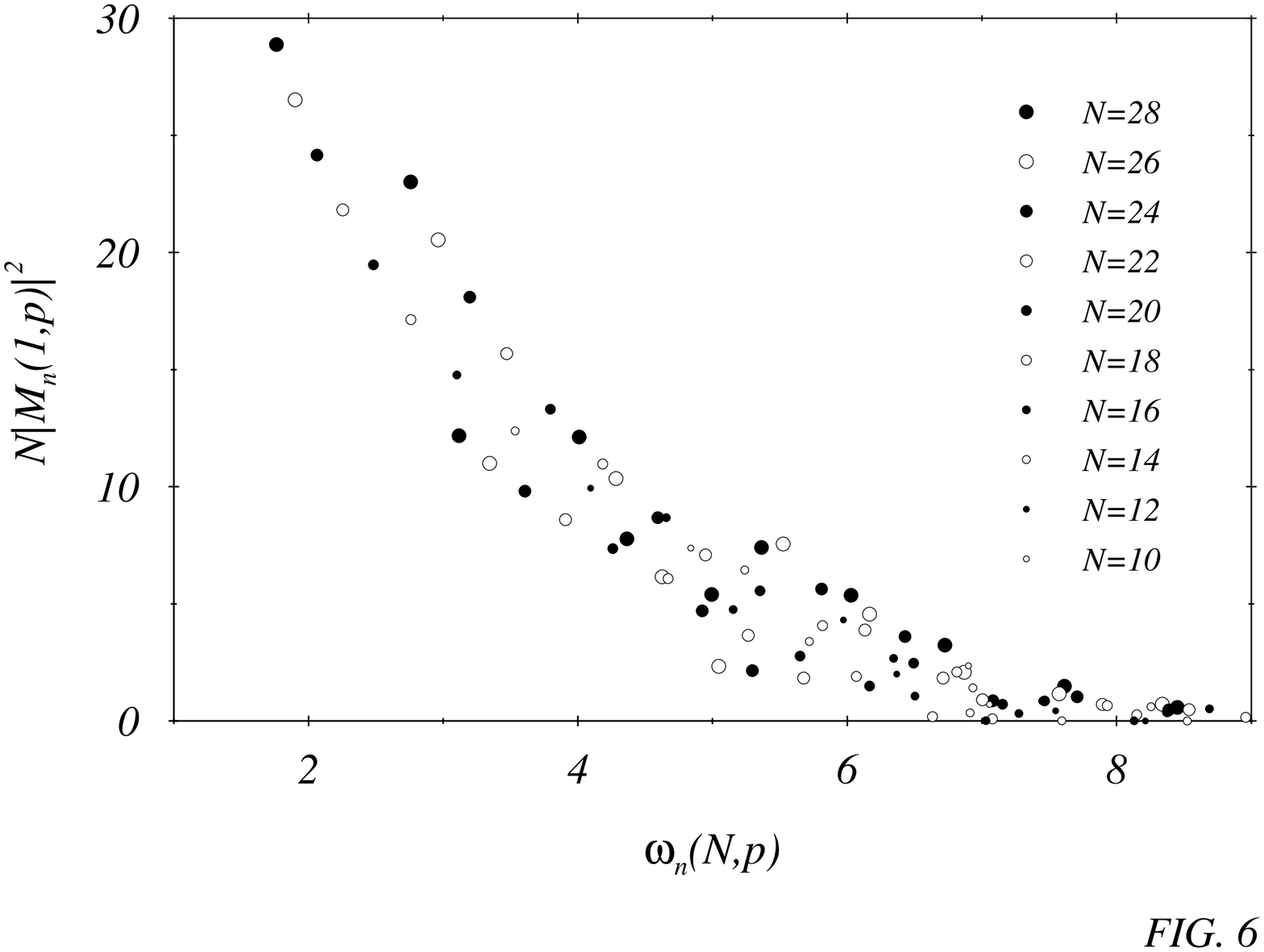,width=6.8cm,angle=-90}}
   \caption{The scaled transition probabilities (\ref{Mm}) for finite systems
     ($N=10,...,28$) vs. the excitation energy.}
   \label{fig:S_w_pi}
\end{figure}

The divergence (\ref{sigma_N_infinity}) of the susceptibility
(\ref{susz_pi_N}) demands for an infrared singularity in the dynamic structure
factor at $p=\pi$
\begin{equation}
  \label{S_w_pi_N}
    S(\omega,\pi,N) \stackrel{\omega\to 0}{\longrightarrow}
    A\frac{\Omega_1}{\omega}\ln\frac{1}{\omega},
\end{equation}
which is visible in the numerical data for the scaled transition probabilities
(\ref{Mm}) of Fig.~\ref{fig:S_w_pi}, in spite of the rather large finite-size
effects.

Insertion of (\ref{S_w_pi_N}) into (\ref{susz_pi_N}) leads to
(\ref{sigma_N_infinity}).

\begin{figure}[ht]
  \centerline{\epsfig{file=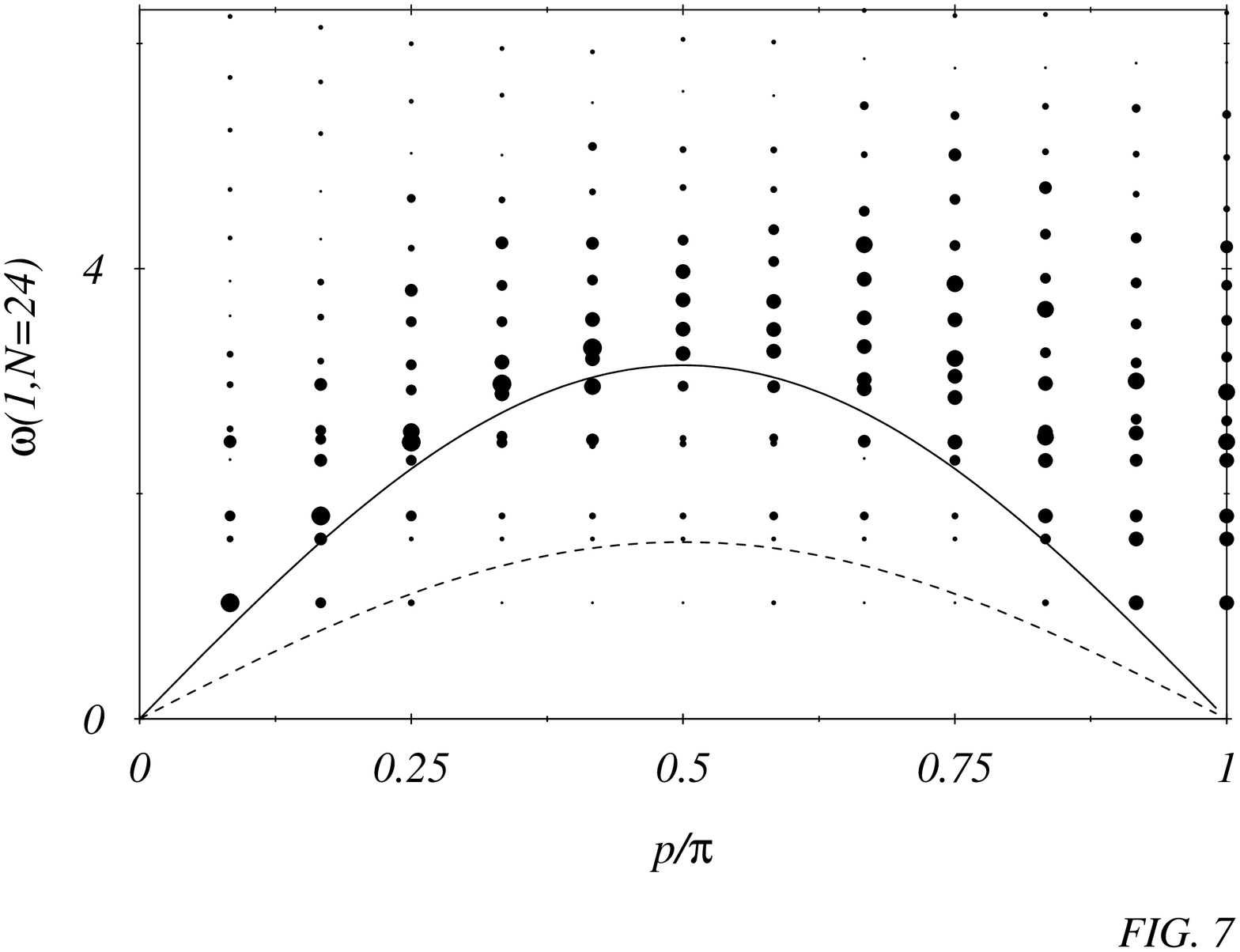,width=6.8cm,angle=-90}}
   \caption{The excitation spectrum for $\theta=1$ and $N=24$. The size of the
     dots is proportional to the relative spectral weights of associated
     transition matrix elements $M_n(1,p)^2$. }
   \label{fig:w_p_N}
\end{figure}

In Fig.~\ref{fig:w_p_N} we present the $p$-dependence of the excitation
energies $\omega_n=E_n-E_0$ for $N=24$ together with the associated relative
spectral weights in percentage terms. Highest spectral weights occur at
excitation energies $\omega_{max}(p)$. This curve is rather well approximated by
\begin{equation}
  \label{w_max}
  \omega_{max}(p)=\pi\sin p.
\end{equation}
The dashed curve is the dispersion relation for the lowest excitations in the
$S_{T}=0$-sector in the thermodynamic limit:
\begin{equation}
  \label{w_L}
  \omega_{L}(p)=\frac{\pi}{2}\sin p,
\end{equation}
as it was calculated in Ref.~\onlinecite{JM72}.  On our finite system we find
nonvanishing excitation also below $\omega_{L}(p)$, they will not survive in the
thermodynamic limit.

Finally let us discuss the connected correlators:
\begin{eqnarray}
  \label{C_p_N}
  C(p,N) &\equiv& 
  \langle \Psi_0(1) | {\bf S}(p)^2 {\bf S}(p)^2 | \Psi_0(1) \rangle 
  \nonumber \\ && -
  \langle \Psi_0(1) | {\bf S}(p)^2 | \Psi_0(1) \rangle^2,
  \nonumber \\
  &=&  
  \int_{\omega_1(N)}^{\infty} d \omega  S(\omega,p,N), 
\end{eqnarray}
which follow by integrating up the dynamic structure factor (\ref{S_w_pi_N}).
The infrared singularity (\ref{S_w_pi_N}) induces a divergence in $C(\pi,N)$ for
$N\to\infty$:
\begin{equation}
  \label{C_pi_N}
   C(\pi,N) \stackrel{N\to \infty}{\longrightarrow} 
  \frac{A}{2}\Omega_1 (\ln N)^2.
\end{equation}

\begin{figure}[ht]
  \centerline{\epsfig{file=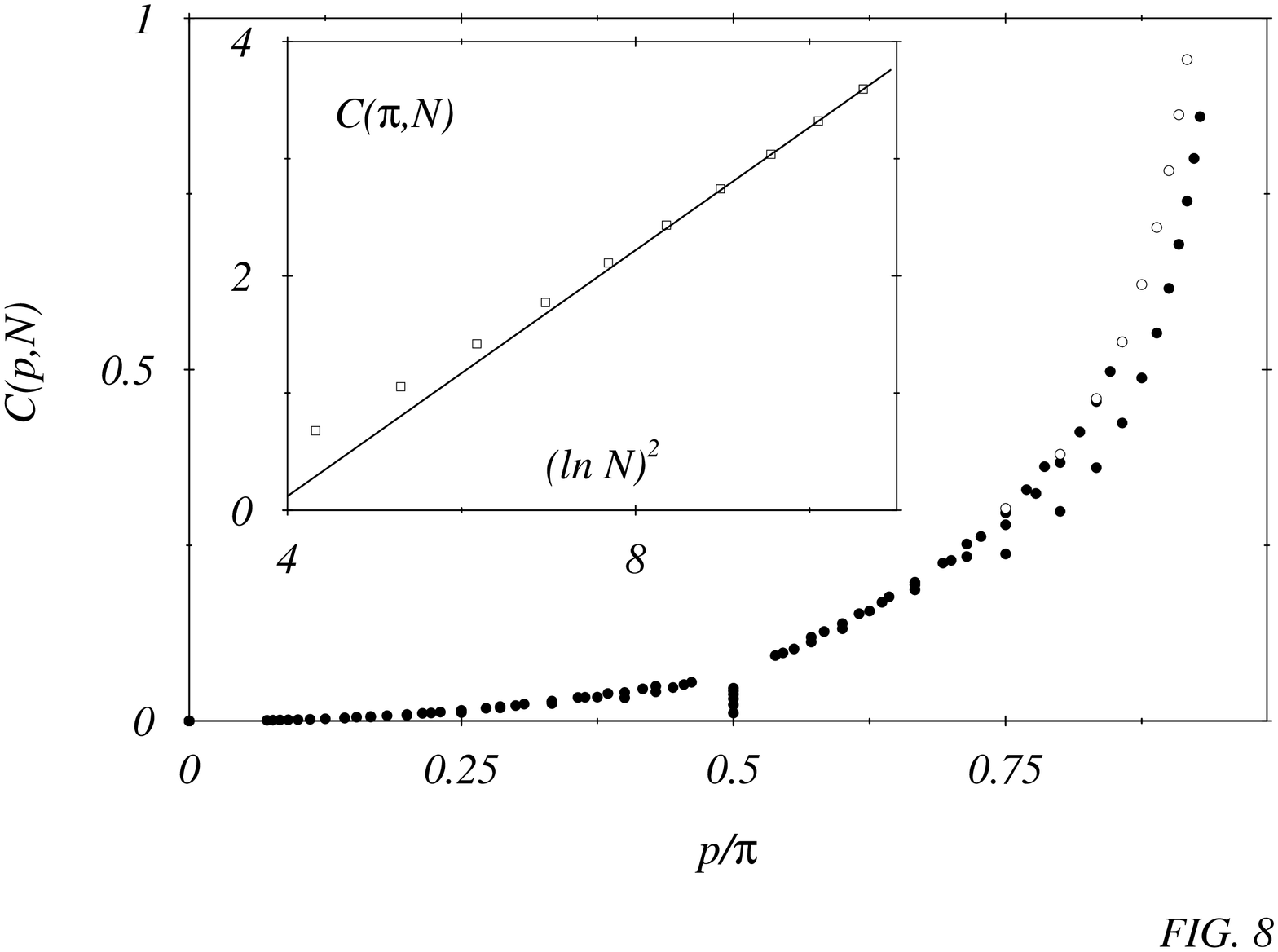,width=6.8cm,angle=-90}}
   \caption{The connected correlators (\ref{C_p_N}) for $N=8,\ldots,28$ are
     shown as solid symbols $(\bullet)$, whereas the prediction of a finite-size
     scaling ansatz is marked by open symbols $(\circ)$. The inset
     represents the singular behavior of $C(\pi,N)$ together with the
     corresponding fit (\ref{C_pi_N}).}
   \label{fig:C_p_N}
\end{figure}

This behavior is clearly seen in Fig.~\ref{fig:C_p_N}. The slope of the
numerical data in the inset of Fig.~\ref{fig:C_p_N} is 0.525(8), which deviates
from the prediction (\ref{AN0}) and (\ref{w1}):
\begin{equation}
  \label{W1}
  \frac{A}{2}\Omega_1 = 0.682(5).
\end{equation}
by about -23\%. 

The $p$-dependence of the static structure factor (\ref{C_p_N}) is given in
Fig.~\ref{fig:C_p_N}. One observes a singularity at $p=\pi$. Finite-size effects
are small, away from the singularity. Near the singularity they can be described
with a finite-size scaling ansatz in the fashion (\ref{Sigma_N_G}) we used for
$\Sigma_1(\theta=1,\pi,p_1,N)$ in Sec.~III.

In summary: The $N$-dependence of the susceptibility
$\Sigma(\theta=1,\pi,\pi,N)$ and of the connected correlators $C(\pi,N)$ as well
as the $p$-dependence of $\Sigma(\theta=1,p,p,N)$ and $C(p,N)$ support the
existence of the infrared singularity (\ref{S_w_pi_N}) in the dynamic structure
factor (\ref{S_w_p_N}).

%
\section{The 1D
  antiferromagnetic Heisenberg model with next to nearest-neighbor couplings}
%
The approach developed in Secs. II - IV is applicable to all Hamiltonians ${\bf
  H}(\vartheta)$ which depend on a parameter $\vartheta$ and reduce to the
1D Heisenberg Hamiltonian for the specific value $\vartheta=1$. As
a further example we consider the Hamiltonian
\begin{equation}
  \label{Hnn}
  {\bf H}(\vartheta) = {\bf H}_1 + (\vartheta-1){\bf H}_2,
\end{equation}
with next to nearest-neighbor couplings
\begin{equation}
  \label{H2}
  {\bf H}_{2} \equiv \sum_{n=1}^N {\bf S}_n \cdot {\bf S}_{n+2}. 
\end{equation}
These couplings strengthen the antiferromagnetism for $\vartheta<1$, but
frustrate the system for $\vartheta>1$. Therefore, the question arises, what
happens here with the staggered magnetization for $\vartheta\leq 1$. Repeating
the arguments which lead to (\ref{dm_dtheta_int}), yields for the first
derivative of the staggered magnetization squared:
\begin{eqnarray}
  \label{dm_dtheta_nn}
  \left.\frac{{d m^\dag(\vartheta)}^2}{d\vartheta}\right|_{\vartheta=1} &=&
  -2\int_0^{\pi(1-2/N)} \frac{dp}{\pi}[ \cos (2p) + \cos p] 
  \nonumber \\ && \times
     \Sigma(\vartheta=1,p,\pi,N).  
\end{eqnarray}
Again $\Sigma(\vartheta=1,p,\pi,N)$ is given by (\ref{sigma}) in terms of the
excitation energies $\omega_m(\vartheta=1)=E_m(\vartheta=1)-E_0(\vartheta=1)$
and transition probabilities $M_m(1,p)$ of the 1D nearest-neighbor
model with Hamiltonian ${\bf H}(\vartheta=1)$. However, insertion of the pole
term contribution (\ref{sigma_p_pi}) leads here to a {\sl finite} contribution
for $dm^\dag(\vartheta)^2/d\vartheta|_{\vartheta=1}$ in the thermodynamic limit.
It is worthwhile to note how the range of the spin-spin couplings -- over 2
lattice spacings in (\ref{Hnn}) and over $k=\sqrt{N\mp 1}$ lattice spacings in
(\ref{H}) -- enters via the Fourier factors $\cos(2p)$ and $\cos(kp)$ in the
formulas (\ref{dm_dtheta_nn}) and (\ref{dm_dtheta_int}) for
$dm^\dag(\vartheta)^2/d\vartheta|_{\vartheta=1}$, respectively. These Fourier
factors generate the different behavior of
$dm^\dag(\vartheta)^2/d\vartheta|_{\vartheta=1}$ in the two cases (\ref{Hnn})
and (\ref{H}).
%
\section{Summary and conclusions}
%
It is convenient to study the transition from one to two dimensions by means of
an interpolating Hamiltonian ${\bf H}(\theta)$ [cf.(\ref{H})] with a parameter
$\theta$, which controls the anisotropy between the horizontal and vertical
nearest-neighbor couplings. The quantity of interest is the staggered
magnetization $m^\dag(\theta)$, which on one hand is supposed to be nonzero for
the anisotropic 2D model ($0\leq\theta<1$) and which on the
other hand is known to be zero in the 1D limit $\theta=1$. In this
paper we have studied the first derivative $dm^\dag(\theta)^2/d\theta$. It can
be expressed in terms of excitation energies $\omega_n=E_n-E_0$ and transition
matrix elements $\langle \Psi_m | {\bf S}(-p){\bf S}(p) | \Psi_0 \rangle$
associated with the operator ${\bf S}(-p)\cdot{\bf S}(p)$. 

In other words, the $\theta$-evolution of the staggered magnetization
(\ref{dm_dtheta_int}) is determined by the dynamics in the sector with total
spin $S_T=0$ and momentum transfer $\Delta p=0$. Energy conservation sum rules
(\ref{energy_conservation}) yield important constraints on the transition
matrix elements $\langle \Psi_m(\theta) |{\bf S}(-p){\bf S}(p) |
\Psi_0(\theta) \rangle$. They turned out to be very helpful in our numerical
analysis of the low-energy excitations at $\theta=1$:
\begin{enumerate}
\item The dynamic structure factor (\ref{S_w_p_N}) has an infrared singularity
  (\ref{S_w_pi_N}) at $p=\pi$. For $0<p<\pi$, there is  a dispersion relation
  $\omega=\omega_{L}(p)$ (\ref{w_L}) for the lowest excitations (with $\Delta
  S=0$ and $\Delta p=0$) in the thermodynamic limit. On finite systems, we find
  excitations below $\omega_L(p)$.
  
\item The existence of the infrared singularity is supported by the
  divergences, which can be clearly seen in the $N$ dependence of the connected
  correlators $C(\pi,N)$  (\ref{C_pi_N}) and the susceptibility
  $\Sigma(1,\pi,\pi,N)$ (\ref{sigma_N_infinity}) and in the $p$-dependence
  of $C(p,N)$ and $\Sigma(1,p,p,N)$ for $p\to\pi$.
 
\item Owing to the energy conservation sum rule (\ref{sigma_sumrule}), the
  divergence in the susceptibility induces a pole (\ref{sigma_p_pi}) in
  $\Sigma(1,p,\pi)$. This pole term fixes the leading large-$N$ behavior
  (\ref{dmdtheta_1}) of the first derivative in the staggered magnetization
  squared at $\theta=1$.
\end{enumerate}

We have considered in this paper the transition from one to two dimensions in
the spin 1/2-antiferromagnetic Heisenberg model only. However, the general
formulas (\ref{dmdtheta}) and sum rule (\ref{energy_conservation}), which
describe the $\theta$-evolution, hold as well in the spin-1 case, where the 1D
model ($\theta=1$) develops the celebrated {\it Haldane
  gap}.\cite{Hald83,Affl89} The existence of this gap already tell us, that
there is no infrared singularity of the type (\ref{susz_pi_N}). Therefore, we do
not expect a divergence (\ref{dmdtheta_1}) in
$dm^{\dag}(\theta)^{2}/d\theta|_{\theta=1}$ in the spin 1 case
%
\begin{appendix}
%
%
\section*{Recursion method}
%
The recursion method is designed to approximately determine the excitation
energies and matrix elements of transition operators, the dynamics of which are
to be studied. Neglecting indices like momentum $p$ and chain length $N$ let us
consider
\begin{equation}
  f_A(\omega) =  \sum_n |\langle n|A|0\rangle |^2\delta[\omega-(E_n-E_0)],
\end{equation}
with $|0\rangle,\,|n\rangle$ being the ground state and the excited states of
the system defined by the Hamiltonian $H$. $E_0$ and $E_n$ are the
corresponding energies. (e.g. A might be chosen as $A\,=\,S_j(p,N),\quad
j=1,2,3$). The Laplace transform of $f_A(\omega)$ reads
\begin{eqnarray}
  f_A(\tau) 
  & = & \sum_n e^{-\omega_n\tau} |\langle n|A|0\rangle|^2,\\
  & = & \langle 0|A^+e^{-({\bf H}-E_0)\tau}A|0\rangle=\langle f_0|f(\tau)\rangle,
\end{eqnarray}
with $|f_0\rangle\,=\,A|0\rangle$ and $|f(\tau)\rangle$ fulfilling
\begin{equation}\label{A:diff_eq}
  \frac{\partial}{\partial \tau}|f(\tau)\rangle =  -\bar{{\bf H}}|f(\tau)\rangle,
  \qquad(\bar{{\bf H}}\,=\,{\bf H}-E_0).
\end{equation}
Now, a Gram-Schmidt construction is used (see also Ref. \onlinecite{VM94}) to
form an orthogonal set of states $\{|f_k\rangle\}$, namely
\begin{equation}\label{A:fkp1}
  |f_{k+1}\rangle =  
  \bar{{\bf H}}|f_k\rangle - a_k|f_k\rangle -b_k^2|f_{k-1}\rangle
\end{equation}
\begin{mathletters}
\begin{eqnarray}
  a_k &=& 
  \frac{\langle f_k|\bar{{\bf H}}|f_k\rangle}{\langle f_k|f_k\rangle},\quad
  k = 0,1,2,..,L-1,\\
  b_0^2 &\equiv&  0, \\
  b_k^2 &=& \frac{\langle f_k|f_k\rangle}{\langle f_{k-1}|f_{k-1}\rangle},\quad 
  k=1,2,..,L-1,
\end{eqnarray}
\end{mathletters}
and $|f(\tau)\rangle$ is expressed as
\begin{equation}
|f(\tau)\rangle =  \sum_{k=0}^{L-1} {\bf D}_k(\tau)|f_k\rangle.
\end{equation}
$L$ denotes the dimension of the Hilbert space, i.e. the number of states
$|n\rangle$ leading to nonzero matrix elements $\langle n|A|0\rangle$. The
differential equation (\ref{A:diff_eq}) leads to
\begin{equation}
  \sum_k \dot{\bf D}_k(\tau)|f_k\rangle =  -\sum_k {\bf D}_k(\tau)\left(
  |f_{k+1}\rangle+a_k|f_k\rangle+b_k^2|f_{k-1}\rangle\right)
\end{equation}
or in matrix notation $\dot{\vec{\bf D}}=-{\bf M \vec{D}}$ with ${\bf M}$ being
a tridiagonal $L\times L$-matrix with eigenvalues $\omega_n$ and eigenvectors
$\vec{e}_n$.  The eigensolutions $v_k(\tau)$ of the set of linear differential
equations read $v_k(\tau)=v_k^0e^{-\omega_k\tau}$ leading to
\begin{equation}
  {\bf D}_k(\tau) =  \sum_n (\vec{e}_n)_kv_n^0 e^{-\omega_n\tau}.
\end{equation}
Here, the $v_k^0$ are given by the initial condition $\vec{\bf D}(\tau=0)\!=\!
(1,0,0,...,0)^T$ yielding for $f_A(\tau)$
\begin{equation}
  f_A(\tau)={\bf D}_0(\tau)\langle f_0|f_0\rangle =  
             \sum_n e^{-\omega_n\tau}v_n^0(\vec{e}_n)_0
             \langle f_0|f_0\rangle,
\end{equation}
i.e. $|\langle n|A|0\rangle|^2=v_n^0(\vec{e}_n)_0\langle f_0|f_0\rangle$.  It
should be noted that the eigenvalues of ${\bf M}$ are real, however, its non
non-symmetric structure does not require the $\vec{e}_n$ to be orthogonal.  The
approximation in our scheme sets in in the fact that the numerical treatment
does not allow for generating the whole set of $L$ orthogonal states. The number
of iterations in (\ref{A:fkp1}) is reduced to $\tilde{L}$, the matrix ${\bf M}$
truncated to $\tilde{L}\times\tilde{L}$ and the evaluation given above is
performed on this reduced set of states. Details and checks for this method are
given in Refs.~\onlinecite{FKMW95,FGMSK96}.

In the present paper the recursion method has been extended [see e.g.  Eq.
(\ref{dmdtheta})] to treat matrix elements of type
\begin{equation}
  \langle g|f(\tau)\rangle =  \langle g|e^{-\tau\bar{{\bf H}}}|f_0\rangle,
\end{equation}
with $|f_0\rangle={\bf A}|0\rangle$, $|g\rangle={\bf B}|0\rangle$, ${\bf A}$
and ${\bf B}$ acting on the same space.  Using the series representation of
$|f(\tau)\rangle$ we obtain
\begin{equation}
  \langle 0|{\bf B}^+|n\rangle\langle n|{\bf A}|0\rangle  =  
  \sum_k v_n^0(\vec{e}_n)_k\langle g|f_k\rangle,
\end{equation}
which contains the transition probabilities $|\langle n|{\bf A}|0\rangle|^2$
for the choice ${\bf B}={\bf A}$.

\end{appendix}
%
%

%
%

\end{document}